\documentstyle[preprint,eqsecnum,aps,epsf]{revtex}
\def\be{\begin{equation}}
\def\ee{\end{equation}}
\def\bearr{\begin{eqnarray}}
\def\eearr{\end{eqnarray}}
\begin{document}
\draft
\preprint{}

\title{MESON CORRELATORS  IN QCD VACUUM - IS SATURATION
THE RIGHT APPROACH ?}
\author{Varun Sheel\footnote{Electronic address:
varun@prl.ernet.in}, Hiranmaya Mishra\footnote{Electronic address:
hm@prl.ernet.in} and Jitendra C. Parikh \footnote{Electronic address:
parikh@prl.ernet.in} }
\address{Theory Group,
Physical Research Laboratory, Navrangpura, Ahmedabad 380 009, India}
\maketitle
\begin{abstract}
Equal time, point to point correlation functions for spatially
separated meson currents are calculated with respect to a
variational construct  for the ground state of QCD. With exact
calculations,
vector, axial vector and scalar channels show qualitative
agreement with the phenomenological predictions, where as
pseudoscalar channel does not. However, the
pseudoscalar correlator, when approximated by saturating
with intermediate one  pion states
agrees with results obtained from spectral density
functions parametrised by pion decay constant and
$<\bar \psi \psi>$
value obtained from chiral perturbation theory.
\end{abstract}
\vskip 0.5cm
\narrowtext

\section{Introduction}

Quantum Chromodynamics (QCD) in the low energy sector is
nonperturbative and the vacuum structure here is
nontrivial \cite{suryak}.
The vacuum structure of QCD has been studied since quite some time
both with quark condensates associated with chiral symmetry breaking
\cite{nambu}  as well as with gluon
condensates \cite{shut,hans}. An interesting quantity to study with
such a nontrivial structure of vacuum is the behaviour of
current-current
correlators illustrating different physics involved at
different spatial distances. This has recently been emphasized
in a review by Shuryak \cite{surcor} and studied through
lattice simulations \cite{neglecor}. The basic point is that
the correlators can be used to study the interquark
interaction --- its dependence on distance. In fact
they complement bound state hadron properties in the same way
that scattering phase shifts provide information about the
nucleon-nucleon force complementary to that provided by the
properties of the deuteron \cite{neglecor}.

We have recently considered the structure of QCD vacuum with
both quark and gluon condensates using a variational ansatz
\cite{prliop}. We shall use here such an explicit construct of
QCD vacuum obtained through energy minimisation \cite{prliop}
to evaluate the meson correlators.

We organise the paper as follows. In section II we
recapitulate the results of Ref.\ \cite{prliop}. In section III we
define and calculate meson correlation functions. In section
IV we quote the results. Section V is devoted to the study of
the exceptional case of pseudoscalar correlator. Finally we
discuss the results in section VI.

\section{QCD VACUUM WITH QUARK AND GLUON CONDENSATES}

We have considered the vacuum structure in QCD using a variational
approach with both quark and gluon condensates \cite{prliop}.
Here we shall very briefly recapitulate the results of the
same for the sake of completeness.
The trial variational ansatz for the QCD vacuum is taken as

\be
|vac>=U_GU_F|0>
\label{vacp}
\ee
obtained through the unitary operators  $U_G$  and  $U_F$ for
gluons and quarks respectively on  the perturbative vacuum $|0>$.

For the gluon sector, the unitary operator $U_G$ is of the form
 \be
U_G
=\exp{({B_G}^\dagger-B_G)}
\ee
 with
the gluon pair creation operator ${B_G}^\dagger $
 given by
\be
{{B_G}^\dagger}={1\over 2}
\int {f(\vec k){{{a^a}_{i}(\vec k)}^\dagger}
{{{a^a}_{i}(-\vec k)}^{\dagger}}d\vec k}
\ee
\noindent
In the above $ {{a^a}_{i}(\vec k)}^\dagger $
are the transverse gluon field creation operators satisfying
the following quantum algebra in Coulumb gauge \cite{prliop}

\be
\left[ {a^a}_{i}(\vec k),{{a^b}_{j}(\vec k^{'})}^\dagger\right]=
 \delta^{ab}({\delta_{ij}}-{k_{i}k_{j}\over k^2})
\delta({\vec k}-{\vec k^{'}})
\ee
with ${a^a}_{i}(\vec k)$ annihilating the perturbative vacuum
$|0>$.
Further $f(k)$ is a trial function associated with
gluon condensates.

Similarly for the quark sector we have,
 \be
U_F
=\exp{({B_F}^{\dagger}-B_F)}
\ee
 with
 \be
 {B_F}^{\dagger}=
\int \bigg[h(\vec k){{c^i}_{I}(\vec k)}^{\dagger}
(\vec \sigma \cdot \hat k)
{\tilde c}^i_{I}(-\vec k)
\bigg]\; d\vec k
\label{bfbeta}
\ee
Here $h(\vec k)$ is a trial function associated
with quark antiquark condensates. The operators
$c^\dagger$ and $\tilde c$ create a quark and antiquark
respectively
when operating on the perturbative vacuum. They satisfy the
anticommutation relations

\be
 [c^{i}_{Ir}(\vec k),c^{j}_{Is}(\vec k')^\dagger]_{+}=
 \delta _{rs}\delta^{ij}\delta(\vec k-\vec k')=
[\tilde c_{Ir}^{i}(\vec k),\tilde c_{Is}^{j}(\vec k')^\dagger]_{+}
\ee

Clearly such a structure for the vacuum eventually
reduces to a Bogoliubov transformation for the operators.
One can then calculate the energy density functional given as

\be
\epsilon_{0} \equiv F(h(\vec k),f(\vec k))
\ee

The condensate functions $f(\vec k)$ and $h(\vec k)$ are to be
determined such that the energy density $\epsilon _0$ is a minimum.
Since the functions cannot be determined analytically through
functional minimisation except for a few  simple cases
\cite{grnv}, we choose
the alternative approach of parameterising the condensate functions as
( with $k=|\vec k|$),
\be
\sinh f(\vec k)=Ae^{-Bk^2/2}
\ee
This corresponds to taking a Gaussian distribution for the
perturbative gluons in the nonperturbative vacuum.
Similarly, for the function $h(\vec k)$ describing the quark antiquark
condensates we take the ansatz,
\be
\tan 2h(\vec k)=\frac{A'}{(e^{R^2 k^2} - 1 )^{1/2}}
\label{ansatz}
\ee
\noindent
Further one could relate the quark condensate function to the
wave function of pion as a quark antiquark bound state
\cite{misra} and hence to the decay constant of pion.

In Ref.\ \cite{prliop} the energy density is minimised
with respect to the condensate parameters
subjected to  the constraints that the pion decay constant
f$_\pi$ and the gluon condensate value
$\frac{\alpha_s}{\pi}<G^a_{\mu\nu}{G^a}^{\mu\nu}>$
 of
Shifman Vainshtein and Zhakarov \cite{svz} come out as the
experimental value of 93 MeV and 0.012 $GeV^4$ respectively.

The results of such a minimisation showed the instability of
the perturbative vacuum to formation of quark antiquark as
well as gluon
condensates when the coupling became greater than 0.6. Further
the charge radius for the pion comes out correctly
($R_{ch}\simeq 0.65$ fm) for $\alpha_s =
1.28$. The corresponding values of A$^{'}$ and R of
Eq.\ (\ref{ansatz}) are calculated to be
$A'_{min}\simeq 1$ and  $R\simeq 0.96$ fm.

With the structure of QCD vacuum thus fixed from pionic
properties  and SVZ value we consider the  meson correlators
in the next section.

\section{MESON CORRELATION FUNCTIONS}

Consider a generic meson current of the form
\be
J(x) = \bar \psi_{\alpha}^{i}(x) \Gamma_{\alpha \beta }
\psi_{\beta}^{j}(x)
\label{current}
\ee
\noindent where
$ x \; $is a four vector;
$\alpha$ and $\beta$ are spinor indices;
i and j are flavour indices;
$\Gamma$ is a $4 \times 4$ matrix $(1, \gamma_{5},
\gamma_{\mu}$  or $\gamma_{\mu} \gamma_{5} )$

Because of the homogeneity of the vacuum we define the conjugate
current to the above at the origin as,
\be
\bar J(0) = \bar \psi_{\lambda}^{j}(0) \Gamma_{\lambda \delta }^{'}
\psi_{\delta}^{i}(0)
\label{currentbar}
\ee
with $\Gamma^{'} = \gamma_{0} \Gamma^{\dagger } \gamma_{0} $

The meson correlation function for the above currents is defined
as,
\be
R(x) = < T J(x) \bar J(0) >_{vac}
\label{cordef}
\ee

 From now on we assume that expectation values are always with
respect to the nonperturbative vacuum of our model,
hence we drop the subscript $vac$.

Hence with Eqs.\ (\ref{current}), (\ref{currentbar})
and (\ref{cordef}) we have
\be
R(x) = \Gamma_{\alpha \beta } \Gamma_{\lambda \delta }^{'}
 < T \bar \psi_{\alpha}^{i}(x)
\psi_{\beta}^{j}(x) \bar \psi_{\lambda}^{j}(0)
\psi_{\delta}^{i}(0) >
\label{cordef1}
\ee
This reduces to the identity
\be
R(x)  =  \Gamma_{\alpha \beta} \Gamma_{\lambda \delta}^{'}
            < T \psi_{\beta}^{j}(x) \bar \psi_{\lambda}^{j}(0) >
           <T \bar \psi_{\alpha}^{i}(x) \psi_{\delta}^{i}(0) >
\label{cordef2}
\ee

The above definition of $R(x)$ is exact since the four point
function does not contribute.
In fact, in the evaluation of Eq.\ (\ref{cordef1}) we shall
have a sum of two terms. The first is
equivalent to the product of two point functions which is
Eq.\ (\ref {cordef2}).
The second term arises from contraction of operators at the
same spatial point, related to
disconnected diagrams and thus can be discarded.

In Eq.\ (\ref{cordef2}) the first term  can be identified
as the interacting quark propagator
\[ S(x) = < T \psi^{j}(x) \bar \psi^{j}(0) > \]
It can be shown using the CPT invariance of the vacuum
\cite{bdrell} that the second term is given as

\bearr
< T \bar \psi^{i}(x) \psi^{i}(0) > & = & - \gamma_{5} S(x) \gamma_{5}
\\ \nonumber
      & = & - S(-x)
\eearr

Hence the correlation function  of Eq.\ (\ref{cordef1}) becomes
\be
R(x) = - Tr \left[ S(x) \Gamma^{'} S(-x) \Gamma \right]
\label{cordef3}
\ee

Similarly the correlator for massless noninteracting quarks can
be given as

\be
R_{0}(x)  =  - Tr \left[ S_{0}(x) \Gamma^{'} S_{0}(-x) \Gamma \right]
\ee

Our task is now to evaluate the expression (\ref{cordef3})
with the ansatz for QCD vacuum as given in Eq.\ (\ref{ansatz}).
Further we shall be interested in evaluating the equal time
point to point correlation functions. With this in mind we
first calculate the equal time interacting Feynman propagator
in this nonperturbative vacuum given as \cite{mac}

\be
S_{\alpha \beta}(\vec x)   =  \left< \frac{1}{2} \left[
 \psi_{\alpha}^{i}(\vec x),
             \bar \psi_{\beta}^{i}(0) \right] \right>
\ee

Remembering that the expectation value is with respect to the
$|vac>$ as given in Eq.\ (\ref{vacp}), the above
propagator reduces to
\be
S(\vec x) = \frac{1}{2} \frac{1}{(2 \pi)^3}
            \int e^{i\vec k .\vec x }d\vec k
\left[ \sin 2 h(\vec k) - (\vec\gamma \cdot
\hat k)~\cos 2 h(\vec k)\right] \label{corint}
\ee

In evaluating the above we have used the expectation values

\begin{mathletters}
\be
<\psi^{i}_\alpha(\vec x)^{\dagger}
\psi^{j}_\beta(\vec y)>=
(2\pi)^{-3}\delta^{ij}\int
\Big ( \Lambda _-(\vec k)\Big )_{\beta\alpha}
e^{-i\vec k .(\vec x-\vec y)}d\vec k
\label{jpj}
\ee
\be
<\psi^{i}_\alpha(\vec x)
\psi^{j}_\beta(\vec y)^{\dagger}>=
(2\pi)^{-3}\delta^{ij}\int
\Big ( \Lambda _+(\vec k)\Big )_{\alpha \beta}
e^{i\vec k .(\vec x-\vec y)}d\vec k
\label{jjp}
\ee
\end{mathletters}

\noindent where,
\be
\Lambda_{\pm}(\vec  k)=\frac{1}{2}\big
(1 \pm \gamma ^0 \sin 2 h(\vec k) \pm
(\vec\alpha \cdot \hat k)~\cos 2 h(\vec k)\big )
\ee

We next use the condensate function as given in
Eq.\ (\ref{ansatz}) to evaluate Eq.\ (\ref{corint}).
Here we shall take the parameters
$A'=1$ and  $R= 0.96$ fm which give the correct
pionic properties \cite{prliop}.
Then the equal time interacting quark propagator Eq.\ (\ref{corint})
reduces to, with $x=|\vec x|$,

\be
S(\vec x) = - \frac{i}{2 \pi^2} \frac{\vec\gamma \cdot \vec x}{x^4}
    +  \frac{1}{(2 \pi)^{3/2}} \frac{1}{2R^3} e^{-x^2 / (2R^2)}
    - \frac{i}{(2 \pi)^2} \frac{\vec\gamma \cdot \vec x}{x^2} I(x)
\ee

\noindent where

\be
I(x) = \int_{0}^{\infty } \left( \cos k x - \frac{\sin k x}{kx} \right)
   \frac{k e^{-R^2 k^2}}{1+(1-e^{-R^2 k^2})^{1/2}} dk
\label{integral}
\ee

We may further note that when $h(k) = 0$ i.e the condensates
vanish, we recover the free massless propagator
\begin{mathletters}
\bearr
S_{0}(x) & = & - \frac{1}{2} \frac{1}{(2 \pi)^3}
     \int d\vec k \; \vec\gamma \cdot \hat k \; e^{i\vec k .\vec x } \\
           & = & - \frac{i}{2 \pi^2} \frac{\vec\gamma \cdot \vec x}{x^4}
\eearr
\end{mathletters}

Having obtained the propagators, we can calculate the correlation
function, Eq.\ (\ref{cordef3}) for a generic current of the form
 as in Eqs.\ (\ref{current}) and (\ref{currentbar}).
For convenience, we will consider the ratio of the physical
correlation function to that of massless noninteracting quarks

\be
\frac{R(x)}{R_{0}(x)} = \left(1+\frac{1}{2} x^2 I(x) \right)^2
   + \frac{\pi}{8} \frac{x^6}{R^6}  e^{-x^2/R^2}
\frac{x^2 Tr\left[ \Gamma^{'} \Gamma \right]}{ x^{i} x^{j}
   Tr\left[\gamma^{i} \Gamma^{'} \gamma^{j} \Gamma \right]}
\label{gencor}
\ee
which is then evaluated in different channels with the
corresponding Dirac structure for the currents.

\section{Results}
We have studied the above ratio of correlators for four
channels. In each channel we associate the current with a
physical meson having quantum numbers identical to that of the
current. The results are shown in Table~\ref{table1} and in
Fig~\ref{exactfigure}.
We may notice some general features and relationships among
the correlators. The pseudoscalar correlator is always greater
than the scalar correlator and vector correlator is greater
than the axial vector correlator. We may emphasize here that
these relations are rather general in the sense that they do
not depend on the {\em explicit} form of the condensate
function and arise due to
the different Dirac structure of the currents which is reflected
in the generic expression for the correlation functions as in
Eq.\ (\ref{gencor}).
The behaviour of each channel is consistent with that predicted
by phenomenology except in the pseudoscalar case where the ratio
does not go as high as expected from phenomenology.
We examine this in the next section.

\section{Pseudoscalar Channel}
\begin{figure}
\mbox{\hskip -1.0in}\epsfbox{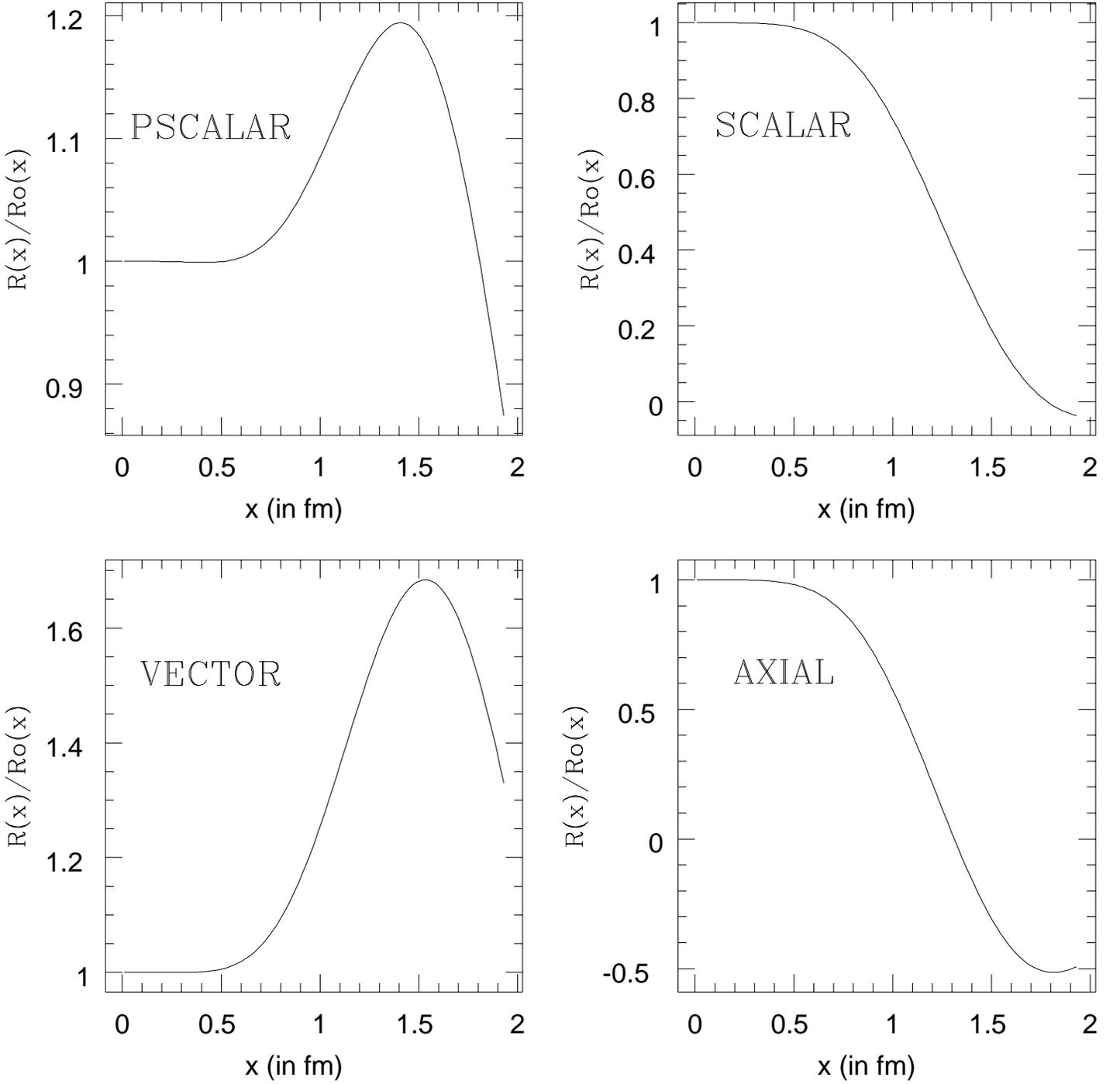}
\caption{The ratio of the meson correlation functions in QCD
vacuum to the
correlation functions for noninteracting massless quarks,
$\displaystyle \frac{R(x)}{R_{0}(x)}  $,
Plotted vs. distance x (in fm) \label{exactfigure}}
\end{figure}

The explicit evaluation of the pseudoscalar correlator gives,
using Eq.\ (\ref{gencor})
\be
\frac{R(x)}{R_{0}(x)} =
  \left[1+\frac{1}{2} x^2 I(x) \right]^2
   + \frac{\pi}{8} \frac{x^6}{R^6} e^{-x^2/R^2}
\ee
which may also be read off from column 4 of Table~\ref{table1}.
This is plotted as a function of $x$ in (Fig.~\ref{exactfigure}).
As may be seen from (Fig.~\ref{exactfigure}) this ratio has a
maximum of $\sim$ 1.2 at $x \sim$ 1.3 fm . Phenomenologically
\cite{surcor} the peak is at $\sim$ 100 at $x \sim$ 0.5. In
order to compare our results with other calculations we
evaluate the same correlator approximately by saturating
intermediate states with one pion states.

Using translational invariance the correlator may be written as

\bearr
R(x) & = & < T J^p(x) \bar J^p(0) > \\
  & = & \frac{1}{2} \left( < J^p(0) \bar J^p(0) >
            e^{i \vec p . \vec x}
   + < \bar J^p(0) J^p(0) >  e^{- i \vec p . \vec x} \right)
\eearr

Also using the fact that for the pseudoscalar current
$J^{p}=\bar u \gamma_{5} d$ and $\bar J^p = -J^p $ we have

\be
R(x)  = \frac{1}{2} < J^p(0) \bar J^p(0) >
   \left(   e^{i \vec p . \vec x} + e^{- i \vec p . \vec x} \right)
\ee

We now insert a complete set of intermediate states between
the two currents but retain only the one pion state
in the sum for the four point function. Thus,

\be
R(x)  = \frac{1}{2} \int <vac \mid  J^p(0)  \mid
        \pi^{a}(\vec p) > < \pi^{a}(\vec p) \mid \bar J^p(0) \mid  vac>
   \left(   e^{i \vec p . \vec x} + e^{- i \vec p . \vec x}
\right) d\vec p
\ee

We may evaluate the above matrix element using the definition
of the pion decay constant given as \cite{lee}
\be
<vac \mid  J^{\mu a}_{5}(x)  \mid \pi^{a}(p) >
= \frac{i f_{\pi} p^{\mu}}{(2\pi)^{3/2} (2p_{0})^{1/2}}
e^{i p \cdot x}
\label{pdecay}
\ee

\noindent where $ J^{\mu a}_{5} = [\bar \psi \gamma^{\mu} \gamma^5
\tau^a \psi ] $ is the axial current.

It can be shown \cite{sakurai} that the divergence
of the axial current gives the pseudoscalar current

\be
\partial_{\mu} [\bar \psi \gamma^{\mu} \gamma^5 \tau^a \psi ] =
2 \; i \;m_q [\bar \psi \gamma^5 \tau^a \psi ]
\label{dividentity}
\ee
where $m_q$ is the current quark mass. Thus
taking divergence of both sides of Eq.\ (\ref{pdecay}) and using Eq.\
(\ref{dividentity}) we get,
\be
2 \; m_q <vac \mid i J^{pa}(x)  \mid \pi^{a}(p) >
     =  \frac{- f_{\pi} m_{\pi}^{2}}{(2\pi)^{3/2} (2p_{0})^{1/2}}
          e^{i p \cdot x}
           \label{divdecay}
\ee

\noindent where we have used $ p^2 = m_{\pi}^2 $ .

In an earlier paper \cite{misra} within our vacuum model
and using the fact that pion is an approximate Goldstone mode
it was demonstrated that
saturating with pion states, gives the familiar current
algebra result

\be
m^{2}_{\pi} = - \frac{m_q}{f_{\pi}^{2}} < \bar \psi \psi >
\ee

With this result we eliminate quark mass $m_q$ in Eq.\ (\ref{divdecay})
in favour of the
quark condensate to get the relation

\be
<vac \mid J^{pa}(\vec x)  \mid \pi^{a}(\vec p) >
=  \frac{i }{2 (2\pi)^{3/2} (2p_{0})^{1/2}}
     \frac{< \bar \psi \psi >}{f_{\pi}}
       e^{i \vec p \cdot \vec x}
\ee

The expression for the pseudoscalar correlator now becomes
\be
R(x)  =  \frac{1}{64 \pi^3} \left( \frac{< \bar \psi
\psi >}{f_{\pi}} \right)^2 \int \frac{1}{(p^2 + m_{\pi}^2 )^{1/2}}
\left(   e^{i \vec p . \vec x} + e^{- i \vec p . \vec x} \right)
d\vec p
\ee

The above integral can be evaluated using the standard integral
\cite{tabintprod}
\[
\int_{0}^{\infty} p(p^2 + \beta^2 )^{\nu - 1/2} sin(\alpha p) dp
= \frac{\beta}{\sqrt{\pi}} \left( \frac{2\beta}{\alpha} \right)^\nu
cos(\nu \pi) \Gamma (\nu + \frac{1}{2}) K_{\nu + 1} (\alpha \beta)
\]
for $\alpha > 0, Re \beta > 0$ and in the limit $\nu \rightarrow
0$. We then finally get  for the correlator
(using saturation of pion states)

\be
R(x) = \frac{1}{16 \pi^2} \left( \frac{< \bar \psi
\psi>}{f_{\pi}} \right)^2 \frac{m_{\pi} K_1(m_{\pi}x)}{x}
\ee
The correlator for free massless quarks as calculated in the
earlier section for pseudoscalar is

\be
R_0(x) = \frac{1}{\pi^4 x^6}
\ee

Hence the ratio is
\be
\frac{R(x)}{R_{0}(x)} = \frac{\pi^2}{16} \left( \frac{< \bar \psi
\psi>}{f_{\pi}} \right)^2  x^5 m_{\pi} K_1(m_{\pi}x)
\ee

\begin{figure}
\vspace{1.5in}
\begin{picture}(0,450)(0,0)
\put(0,-250){\mbox{\hskip -1.2in}\epsfbox{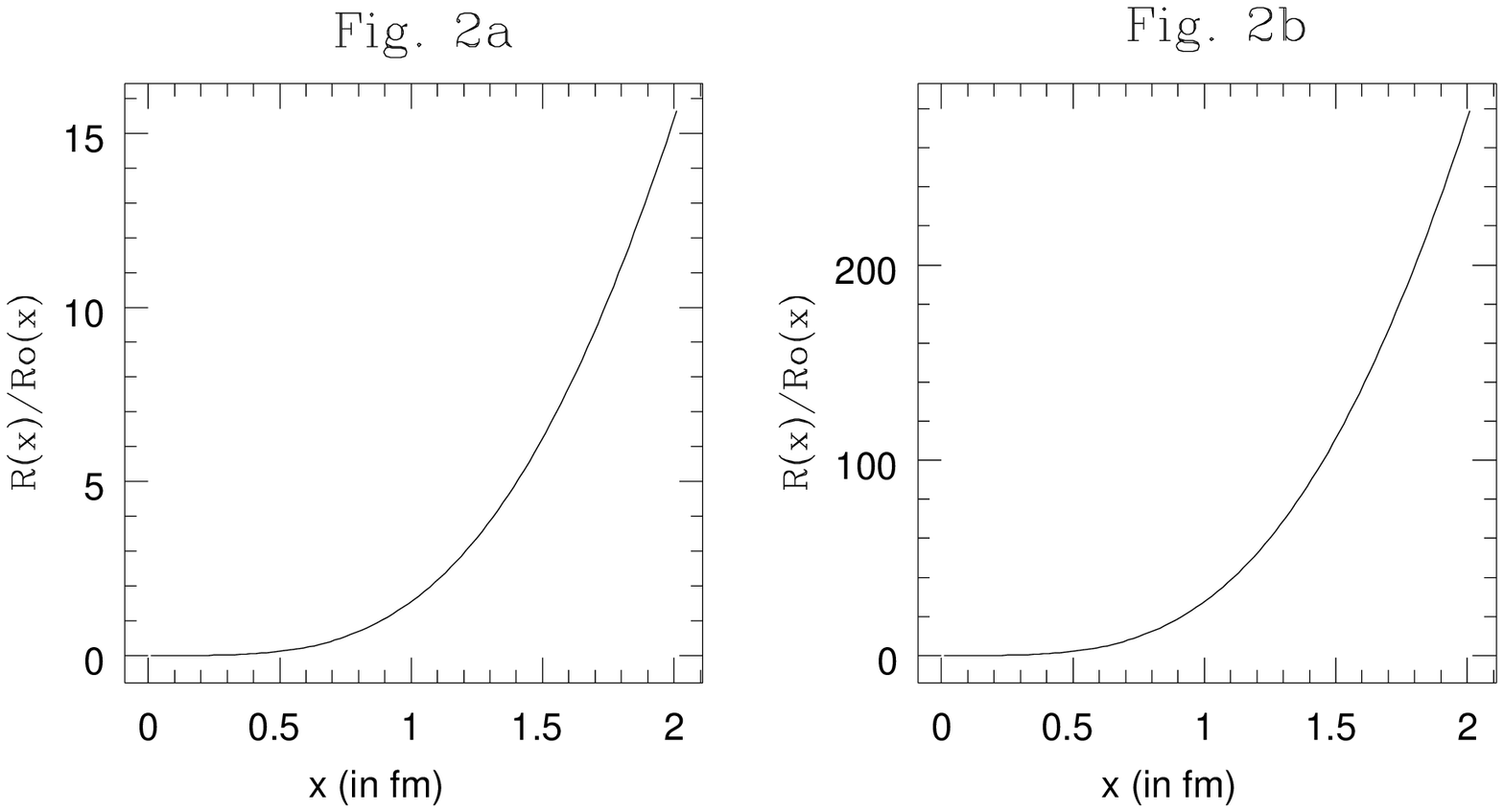}}
%\mbox{\hskip -1.2in}\epsfbox{fig2.ps}
\end{picture}
\caption{The pseudoscalar correlator plotted for  our value
of $<\bar \psi \psi> = $ (190 MeV)$^3$ in (a) and that of Shuryak
%\cite{surcor,surpap}
$<\bar \psi \psi> = $ (307.4 MeV)$^3$  in (b)
 \label{satfigure}}
\end{figure}

We have plotted in Fig.~\ref{satfigure} this ratio for our
value of $<\bar \psi \psi>$ and that used by Shuryak
\cite{surcor,surpap}.
Note that our value of $<\bar \psi \psi >$ is an output of the
variational calculation consistent with low energy hadronic
properties \cite{prliop}.
 We thus observe that the
approximate calculation of the pseudoscalar correlator due to
saturation with one pion states (Fig.~\ref{satfigure}(a)) yields
higher values ($\simeq$ 15 times more) as compared
to the exact calculations (Fig.~\ref{exactfigure}).
Thus the fermionic condensate model for QCD vacuum \cite{prliop}
does not give as high values for the pseudoscalar correlator
as required by phenomenological results.
The value we have used for $< \bar \psi \psi > \simeq$ (190 MeV)$^3$
 is smaller than Shuryak's value of (307.4 MeV)$^3$ \cite{note}
which appears in the parameterisation
of the physical spectral density through the coupling constant
\cite{surpap}. With his value of $< \bar \psi \psi >$ the
ratio $R(x)/R_0(x)$ is shown in (Fig.~\ref{satfigure}(b))
which agrees with phenomenology.

\section{Summary and Discussions}
We have evaluated the mesonic correlators in this paper using a
variational construct for the QCD vacuum. Except for the
pseudoscalar channel the results show qualitative agreement
with phenomenological results \cite{surcor}. Following Shuryak
\cite{surcor,surpap}, we also see that using current algebra
approach the pseudoscalar correlator rises sharply with
spatial separation. Let us recall that the current
algebra result also follows from the approximation of
saturating by one pion states in the normalisation of the pion
state \cite{misra}.

It might appear that by suitably changing the value of
$< \bar \psi \psi >$ in the exact calculation one might be
able to reproduce all the phenomenological results. Actually
we find that it is not so. In fact, it adversely affects the
correlators in the other channel which can be seen in
the exact expressions given in column 4 of
Table~\ref{table1}.

In view of these findings, it is not clear whether saturation
of intermediate states by one pion states only in the
evaluation of the correlator, is sufficiently well justified.
We therefore think that a unified treatment of correlation
functions in all the channels is still not available.

\acknowledgments

VS and HM wish to acknowledge discussions with A. Mishra,
S. P. Misra and P. K. Panda.  We also thank A. Mishra for
a critical reading of the  manuscript.

\def \sur { E.V. Shuryak, {\it The QCD vacuum,
 hadrons and the superdense matter}, (World Scientific,
Singapore, 1988).}

\def \nambu{ Y. Nambu, Phys. Rev. Lett. 4, 380 (1960);
  Y. Nambu and G. Jona-Lasinio, Phys. Rev. 122, 345
(1961); ibid, 124, 246 (1961);
 J.R. Finger and J.E. Mandula, Nucl. Phys. B199, 168 (1982);
 A. Amer, A. Le Yaouanc, L. Oliver, O. Pene and
 J.C. Raynal, Phys. Rev. Lett. 50, 87 (1983);
 ibid, Phys. Rev. D28, 1530 (1983); S.L. Adler and A.C.  Davis,
 Nucl. Phys. B244, 469 (1984); R. Alkofer and P. A.  Amundsen,
 Nucl. Phys.B306, 305 (1988); A.C. Davis and A.M.  Matheson,
 Nucl. Phys. B246, 203 (1984); S. Schramm and W. Greiner,
Int. Jour. Mod. Phys. E1, 73 (1992).}

\def \shut{D. Schutte, Phys. Rev. D31, 810 (1985).}

\def \hans {T. H. Hansson, K. Johnson, C. Peterson,
           Phys. Rev. D26, 2069 (1982).}

\def\surcor {E.V. Shuryak, Rev. Mod. Phys. 65, 1 (1993)}

\def \neglecor{M.-C. Chu, J. M. Grandy, S. Huang and
J. W. Negele, Phys. Rev. D48, 3340 (1993);
ibid, Phys. Rev. D49, 6039 (1994)}

\def \prliop{ A. Mishra, H. Mishra, S.P. Misra, P.K. Panda
and Varun Sheel, preprint hep-ph/9404255}

\def \svz{ M.A. Shifman, A.I. Vainshtein and V.I. Zakharov,
Nucl. Phys. B147, 385, 448 and 519 (1979)}

\def \bdrell { See e.g. in J. D. Bjorken and S. D. Drell,
{\it Relativistic Quantum Fields} (McGraw-Hill, New York, 1965),
Pages 155 and 213}

\def\mac {M. G. Mitchard, A. C. Davis and A. J. Macfarlane,
 Nucl. Phys. B 325, 470 (1989)}

\def \lee { T. D. Lee,
{\it Particle Physics and introduction to Field Theory}
(Harwood Academic, 1982), Page 791 }

\def \sakurai { J. J. Sakurai,
{\it Currents and Mesons} (Chicago Lectures in
Physics, 1969), Page 18 }

\def \grnv { H. Mishra, S.P. Misra and A. Mishra,
 Int. J. Mod. Phys. A3, 2331 (1988);
 S.P. Misra, Phys. Rev. D35, 2607 (1987).}

 \def \misra{A. Mishra, H. Mishra and S. P. Misra, Z. Phys.
C57, 241 (1993) ;
H.Mishra and S.P. Misra, Phys Rev. D48, 5376 (1993) ;
A. Mishra and S.P. Misra, Z. Phys. C58, 325 (1993).}

 \def \tabintprod{{\it Table of Integrals, Series and
Products} edited by I. S. Gradshteyn and I.M. Ryzhik
(Academic Press,1980), Page 427}

\def \note{Our definition of the
condensate value differs from the standard one by a
factor of $2^{1/2}$ }

\def\surpap {E.V. Shuryak, Nucl. Phys. B 319, 541 (1989)}

\begin{table}
\caption{Meson currents and correlation functions \label{table1}}
\begin{tabular}{dddd}
CHANNEL & CURRENT & PARTICLE & CORRELATOR \tablenotemark[1]\\
 & & ($J^{P}$,MASS in MeV) &
$\displaystyle \left[ \; \frac{R(x)}{R_{0}(x)} \; \right]$ \\
 & & & \\
\tableline
%****************************************************
 & & & \\
Pseudoscalar & $J^{p}=\bar u \gamma_{5} d$  & $ \pi^{0} (0^{-},135)$
&  $\left[1+\frac{1}{2} x^2 I(x) \right]^2
   + \frac{\pi}{8} \frac{x^6}{R^6} e^{-x^2/R^2}$\\
 & & & \\
\hline
%****************************************************
 & & & \\
Scalar & $J^{s}=\bar u d$  & $ none (0^{+})$
& $\left[1+\frac{1}{2} x^2 I(x) \right]^2
   - \frac{\pi}{8} \frac{x^6}{R^6} e^{-x^2/R^2}$  \\
 & & & \\
\hline
%****************************************************
 & & & \\
Vector & $J_{\mu}=\bar u \gamma_{\mu} d$  & $ \rho^{\pm} (1^{-},770)$
& $\left[1+\frac{1}{2} x^2 I(x) \right]^2
   + \frac{\pi}{4} \frac{x^6}{R^6} e^{-x^2/R^2}$  \\
 & & & \\
\hline
%****************************************************
 & & & \\
Axial & $J_{\mu}^{5}=\bar u \gamma_{\mu} \gamma_{5} d$
& $ A_{1} (1^{+},1100)$
& $\left[1+\frac{1}{2} x^2 I(x) \right]^2
   - \frac{\pi}{4} \frac{x^6}{R^6} e^{-x^2/R^2}$  \\
 & & & \\
\end{tabular}
\tablenotetext[1]{The integral I(x) is defined in
Eq.\ (\ref{integral})}
\end{table}

\begin{references}
\bibitem{suryak}\sur
\bibitem{nambu}\nambu
\bibitem{shut}\shut
\bibitem{hans}\hans
\bibitem{surcor}\surcor
\bibitem{neglecor}\neglecor
\bibitem{prliop}\prliop
\bibitem{grnv}\grnv
\bibitem{misra}\misra
\bibitem{svz}\svz
\bibitem{bdrell}\bdrell
\bibitem{mac}\mac
\bibitem{lee}\lee
\bibitem{sakurai}\sakurai
\bibitem{tabintprod}\tabintprod
\bibitem{note}\note
\bibitem{surpap}\surpap
\end{references}
\end{document}